# Defining Urban and Rural Regions by Multifractal Spectrums of Urbanization

Yanguang Chen

(Department of Geography, College of Urban and Environmental Sciences, Peking University, Beijing, 100871, China)

**Abstract:** The spatial pattern of urban-rural regional system is associated with the dynamic process of urbanization. How to characterize the urban-rural terrain using quantitative measurement is a difficult problem remaining to be solved. This paper is devoted to defining urban and rural regions using ideas from fractals. A basic postulate is that human geographical systems are of self-similar patterns associated with recursive processes. Then multifractal geometry can be employed to describe or define the urban and rural terrain with the level of urbanization. A space-filling index of urban-rural region based on the generalized correlation dimension is presented to reflect the degree of geo-spatial utilization in terms of urbanization. The census data of America and China are adopted to show how to make empirical analyses of urban-rural multifractals. This work is not so much a positive analysis as a normative study, but it proposes a new way of investigating urban and rural regional systems using fractal theory.

## 1 Introduction

Fractals suggest a kind of optimized structure in nature. A fractal object can occupied its space in the best way. Using the ideas from fractals to design cities as systems or systems of cities will help human being make the most of geographical space. In the age or countries of population explosion and land scarcity, it is significant to develop the theory of fractal cities and the method of fractal planning. Fractal geometry has been employed to research cities for about 30 years, producing many interesting or even important achievements (e.g. Arlinghaus, 1985; Arlinghaus and Arlinghaus, 1989; Batty and Longley, 1994; Benguigui *et al*, 2000; Chen, 2014a; Dendrinos and El Naschie, 1994; Frankhauser, 1994; De Keersmaecker *et al*, 2003; Feng and Chen, 2010; Fotheringham *et al*, 1989; Longley *et al*, 1991; Lu and Tang, 2004; Manrubia *et al*, 1999; Rodin and Rodina, 2000; Sambrook and Voss, 2001; Shen, 2002; Sun and Southworth, 2013; Sun *et al*, 2014; Tannier *et al*, 2011; White and Engelen, 1994; Thomas *et al*, 2010; Thomas *et al*, 2007; Thomas *et al*, 2008; Thomas *et al*, 2012). The basic properties of the previous studies by means of fractal theory are as follows. First, more works were focused on cities, but fewer works were devoted to rural systems or urban-rural regional systems. All cities take root deep in rural hinterland. Rural regions are very meaningful for fractal urban studies (Shan and Chen, 1998). Second, more works were based on the concepts from monofractals, but fewer works were on the basis of the notions from multifractals. Cities and systems of cities in the real world are in fact multifractals with multi-scaling rather than monofractals with single scaling. Multifractal method has been applied to urban form (Ariza-Villaverde *et al*, 2013; Chen and Wang, 2013), regional population (Appleby, 1996; Chen and Shan, 1999), urban and rural settlement including central places and rank-size distributions, and so on (Chen, 2014b; Chen and Zhou, 2004; Haag, 1994; Hu *et al*, 2012 ; Liu and Chen, 2003).

Cities and networks of cities are self-organized complex systems (Allen, 1997; Batty, 2005; Portugali, 2000; Portugali, 2011; Wilson, 2000), and fractal geometry founded by Mandelbrot (1982) is a powerful tool for exploring spatial complexity (Batty, 2008; Frankhauser, 1998). Fractals provide new ways of understanding cities. A city bears a nature of recursion, which is the process of repeating items in a self-similar way. This recursive process results in a hierarchical pattern with cascade structure (Batty and Longley, 1994; Chen, 2012; Frankhauser, 1998; Kaye, 1994). For example, if we go into a city, we can find different sectors including residential sector, commercial-industrial sector, open space, and vacant land; if we go into a sector, say, the commercial-industrial sector, we can see different districts, including residential district, commercial-industrial district, open space, and vacant land; if we go into a district, say, the open space district, we can see different neighbourhoods, including residential neighbourhood, commercial-industrial neighbourhood, open space, and vacant land; if we further go into a neighbourhood, say, the vacant land neighbourhood, we can see different sites, including residential site, commercial-industrial site, open space, and vacant land, and so on (Kaye, 1994). The cascade structure resulting from spatial recursion can be generalized to the whole human geographical systems, including urban and rural terrains. If so, multifractal geometry can be employed to characterize or define urban-rural geographical form and landscape.

Among various urban problems remaining to be solved, spatial characterization of urban-rural terrain is very important but hard to be dealt with. The ideas from multifractals can be used to define the urban and rural region. The patterns of urban-rural terrain are associated with dynamics process of urbanization. The ratio of urban population to total population and urban-rural binary can act as a probability measurement and a spatial scale. Thus a multifractal model can be built in terms of levels of urbanization. The rest parts of this article are organized as follows. In Section 2, a theoretical model of multifractal urban-rural structures will be proposed; In Section 3, empirical analyses will be made according to the levels of urbanization of America and China; In Section 4, several questions will be discussed, and finally, the study will be concluded with a brief summary of the main viewpoints.

# 2 Multifractal model of urban-rural regions

## 2.1 Postulate

First of all, a postulate is put forward that human geographical systems bear recursive processes, which result in self-similar patterns. Thus, multifractal geometry can be employed to model the spatial structure of urban and rural regions. To build the multi-scaling model, we must make clear the concept of urbanization. Urbanization indicates a geographical process of increasing number of people that live in urban areas. It results in the physical growth of cities and evolution of urban systems (Knox and Marston, 2007). The basic and important measurement of urbanization is termed "level of urbanization", which denotes the ratio of urban population to the total population (Karmeshu, 1988; United States, 1980; United States, 2004). The total population in a geographical region ($P$) falls into two parts: urban population ($u$) and rural population ($r$). So the level of urbanization can be defined as $L=u/P$, or $L=u/P*100\%$, where $P=u+r$. In fact, there is no clear borderline between urban regions and rural regions. Therefore, there is no distinct difference between urban population and rural population at the macro level. In this case, the level of urbanization depends on the definition of the urbanized area. Each urbanized area can be distributed into two parts: urban regions and rural regions; each non-urbanized area can also be divided into two parts: rural population and urban population. As a result, the urban areas and rural areas form a hierarchical nesting structure (Table 1).

**Table1 Cascade structure of urban areas and rural areas in a geographical region**

| Class | Population distribution | | | | | | | |
|---|---|---|---|---|---|---|---|---|
| 0 | Regional population $P$: 1 unit | | | | | | | |
| 1 | Urban population ($u$): $L$ | | | | Rural population ($r$)：1-$L$ | | | |
| 2 | Non-agricultural population $L^2$ | | Agricultural population $L(1-L)$ | | Non-agricultural population $(1-L)L$ | | Agricultural population $(1-L)^2$ | |
| 3 | Non-agricultural service population $L^3$ | Agricultural service population $(1-L)L^2$ | Non-agricultural service population $L^2(1-L)$ | Agricultural service population $L(1-L)^2$ | Non-agricultural service population $(1-L)L^2$ | Agricultural service population $(1-L)^2L$ | Non-agricultural service population $L(1-L)^2$ | Agricultural service population $(1-L)^3$ |
| … | … | … | … | … | … | … | … | … |
| $n$ | Binomial distribution:  $C_n^r L^{n-r}(1-L)^r$ , ($r$=1,2,…,$n$) | | | | | | | |

## 2.2 Model

Suppose that the spatial disaggregation complies with the two-scale rule, that is, the formation process of urban and rural pattern is dominated by the probability measures, $L$ and 1-$L$; accordingly, the spatial scale $r$ is based on urban-rural binary, that is, $r$=1/2. We will have a hierarchy of urban and rural regions with cascade structure, and a fractal geographical region will emerge. Using the ideas from multifractals (Feder, 1988; Vicsek, 1989), we can define a $q$-order scaling exponent such as

$$\tau(q) = -\frac{\ln[L^q + (1-L)^q]}{\ln 2},$$ (1)

in which $q$ denotes the order of moment, and $\tau(q)$ is called "mass exponent". Derivative of $\tau(q)$

with respect to $q$ yields a scaling exponent as below:

$$\alpha(q) = \frac{d\tau}{dq} = -\frac{1}{\ln 2} \frac{L^q \ln L + (1-L)^q \ln(1-L)}{L^q + (1-L)^q},$$ (2)

where $\alpha(q)$ is the singularity exponent of multi-scaling fractals. By the Legendre transform (Feder, 1988; Vicsek, 1989), we get a local fractal dimension

$$
\begin{aligned}
f(\alpha) &= q\alpha(q) - \tau(q) \\
&= -\frac{q}{\ln 2} \cdot \frac{L^q \ln L + (1-L)^q \ln(1-L)}{Z^q + (1-Z)^q} + \frac{\ln[L^q + (1-L)^q]}{\ln 2}, \\
&= \frac{1}{\ln 2}[\ln[L^q + (1-L)^q] - \frac{L^q \ln L^q + (1-L)^q \ln(1-L)^q}{L^q + (1-L)^q}]
\end{aligned}
$$ (3)

where $f(\alpha)$ is the local dimension of fractal subsets, corresponding to the sub-regions of urban population or rural population. Further, the generalized correlation dimension can be given by

$$D_q = \begin{cases} \dfrac{\tau(q)}{q-1}, & q \neq 1 \\ -\dfrac{L \ln L + (1-L)\ln(1-L)}{\ln 2}, & q = 1 \end{cases},$$ (4)

or

$$D_q = \frac{1}{q-1}[q\alpha(q) - f(\alpha)] ,$$ (5)

where $D_q$ is the global dimension of a geographical fractal set.

The generalized correlation dimension can be defined by a transcendental equation, which is based on the probability measures $L$ and $1$-$L$ and the spatial scale $r$=1/2. The transcendental equation is as below:

$$L^q \left(\frac{1}{2}\right)^{(1-q)D_q} + (1-L)^q \left(\frac{1}{2}\right)^{(1-q)D_q} = 1 ,$$ (6)

in which $(1-q)D_q = \tau(q)$ for $q \neq 1$. It can be proved that the extreme values of $D_q$ are as follows

$$D_{-\infty} = \begin{cases} \ln(L)/\ln(1/2), & L < 1/2 \\ \ln(1-L)/\ln(1/2), & L > 1/2 \end{cases},$$ (7)

$$D_{+\infty} = \begin{cases} \ln(1-L)/\ln(1/2), & L < 1/2 \\ \ln(L)/\ln(1/2), & L > 1/2 \end{cases}.$$ (8)

Using the two formulae, equations (7) and (8), we can calculate the maximum and minimum values of the generalized correlation dimension, $D_{-\infty}$ and $D_{+\infty}$.

The multifractal parameters can be grouped under two heads: the global parameters and the local parameters. The former includes the generalized correlation dimension $D_q$ and the mass exponent $\tau(q)$, and the latter comprises the singularity exponent $\alpha(q)$ and the local fractal dimension $f(\alpha)$. If urbanization process follows the scaling law, we can characterize the urban and rural patterns using the multifractal parameters. For urbanization, if the rural population in a geographical region decreases, the people in subregions and subsubregions will decrease according to certain proportion. Meanwhile, the urban people in each level of regions will increase in terms of corresponding proportion. If a process of urbanization leads to a multifractal pattern, we can estimate the multifractal spectrums by means of the observational data of the level of urbanization.

There are two approaches to yielding multifractal parameter spectrums: one is empirical approach, and the other, the theoretical approach (Chen, 2014b). The first approach is based on whole sets of observational data, which are obtained by some kind of measurement methods such as the box-counting method (Chen and Wang, 2013; Chen, and Zhou, 2001; Liu and Chen, 2003). The second approach is based on theoretical postulates of scaling process, and the multifractal dimension spectrums can be created with simple probability values and scale ratios (Chen, 2012; Chen and Zhou, 2004). In next section, empirical analyses will be made through the second approach, namely, the theoretical approach.

# 3 Empirical analyses

## 3.1 Multifractal spectrums of the US urban-rural structure

The United States of America (USA) is a well-known developed country and its level of urbanization is more 80% today. According to the US census data, its urbanization level, namely, the ratio of urban population to total population, is about $L$=0.8073; thus the ratio of rural population to total population is around 1-$L$=0.1927. Based on this numbers, the multifractal parameters of US urbanization can be estimated in the theoretical way. Using equations (1) and (2), we can calculate the global parameters, including the generalized correlation dimension $D_q$ and the mass exponent $(q)$; using equations (2) and (3), we can compute the local parameters, including the singularity exponent $\alpha(q)$ and the local fractal dimension $f(\alpha)$. In fact, it is easy to reckon the mass exponent and the singularity exponent using the $L$ value and equations (1) and (2); then by means of Legendre's transform, we can obtain the generalized correlation dimension and the local fractal dimension. The principal results of the multifractal parameter are tabulated as follows (Table 2). The change of the multifractal parameters with the moment order $q$ can be displayed with four curves (Figure 1).

**Table 2 Partial values of the multifractal parameters of the US's and China's urban and rural population distributions based on level of urbanization (2010)**

| $q$ | American urban-rural distribution | | | | Chinese urban-rural distribution | | | |
|---|---|---|---|---|---|---|---|---|
| | $\tau(q)$ | $D_q$ | $\alpha(q)$ | $f(\alpha)$ | $\tau(q)$ | $D_q$ | $\alpha(q)$ | $f(\alpha)$ |
| -400 | -950.2577 | 2.3697 | 2.3756 | 0.0000 | -403.7138 | 1.0068 | 1.0092 | 0.0525 |
| -200 | -475.1288 | 2.3638 | 2.3756 | 0.0000 | -201.9600 | 1.0048 | 1.0079 | 0.3724 |
| -100 | -237.5644 | 2.3521 | 2.3756 | 0.0000 | -101.2802 | 1.0028 | 1.0052 | 0.7557 |
| -50 | -118.7822 | 2.3291 | 2.3756 | 0.0000 | -51.0741 | 1.0015 | 1.0029 | 0.9297 |
| -40 | -95.0258 | 2.3177 | 2.3756 | 0.0000 | -41.0479 | 1.0012 | 1.0023 | 0.9542 |
| -30 | -71.2693 | 2.2990 | 2.3756 | 0.0000 | -31.0273 | 1.0009 | 1.0018 | 0.9739 |
| -20 | -47.5129 | 2.2625 | 2.3756 | 0.0000 | -21.0124 | 1.0006 | 1.0012 | 0.9883 |
| -10 | -23.7564 | 2.1597 | 2.3756 | 0.0000 | -11.0032 | 1.0003 | 1.0006 | 0.9971 |
| -5 | -11.8793 | 1.9799 | 2.3740 | 0.0091 | -6.0009 | 1.0001 | 1.0003 | 0.9993 |
| -4 | -9.5073 | 1.9015 | 2.3690 | 0.0314 | -5.0006 | 1.0001 | 1.0003 | 0.9995 |
| -3 | -7.1464 | 1.7866 | 2.3479 | 0.1027 | -4.0004 | 1.0001 | 1.0002 | 0.9997 |
| -2 | -4.8312 | 1.6104 | 2.2642 | 0.3027 | -3.0002 | 1.0001 | 1.0001 | 0.9999 |
| -1 | -2.6845 | 1.3422 | 1.9774 | 0.7071 | -2.0001 | 1.0000 | 1.0001 | 1.0000 |
| 0 | -1.0000 | 1.0000 | 1.3422 | 1.0000 | -1.0000 | 1.0000 | 1.0000 | 1.0000 |
| 1 | 0.0000 | 0.7071 | 0.7071 | 0.7071 | 0.0000 | 1.0000 | 1.0000 | 1.0000 |
| 2 | 0.5377 | 0.5377 | 0.4202 | 0.3027 | 0.9999 | 0.9999 | 0.9999 | 0.9999 |
| 3 | 0.9069 | 0.4535 | 0.3365 | 0.1027 | 1.9998 | 0.9999 | 0.9999 | 0.9997 |
| 4 | 1.2305 | 0.4102 | 0.3155 | 0.0314 | 2.9996 | 0.9999 | 0.9998 | 0.9995 |
| 5 | 1.5429 | 0.3857 | 0.3104 | 0.0091 | 3.9994 | 0.9999 | 0.9997 | 0.9993 |
| 10 | 3.0881 | 0.3431 | 0.3088 | 0.0000 | 8.9973 | 0.9997 | 0.9994 | 0.9971 |
| 20 | 6.1761 | 0.3251 | 0.3088 | 0.0000 | 18.9888 | 0.9994 | 0.9989 | 0.9883 |
| 30 | 9.2642 | 0.3195 | 0.3088 | 0.0000 | 28.9745 | 0.9991 | 0.9983 | 0.9739 |
| 40 | 12.3522 | 0.3167 | 0.3088 | 0.0000 | 38.9544 | 0.9988 | 0.9977 | 0.9542 |
| 50 | 15.4403 | 0.3151 | 0.3088 | 0.0000 | 48.9288 | 0.9985 | 0.9972 | 0.9297 |
| 100 | 30.8806 | 0.3119 | 0.3088 | 0.0000 | 98.7257 | 0.9972 | 0.9948 | 0.7557 |
| 500 | 154.4029 | 0.3094 | 0.3088 | 0.0000 | 495.3957 | 0.9928 | 0.9908 | 0.0177 |
| 1000 | 308.8058 | 0.3091 | 0.3088 | 0.0000 | 990.7962 | 0.9918 | 0.9908 | 0.0001 |

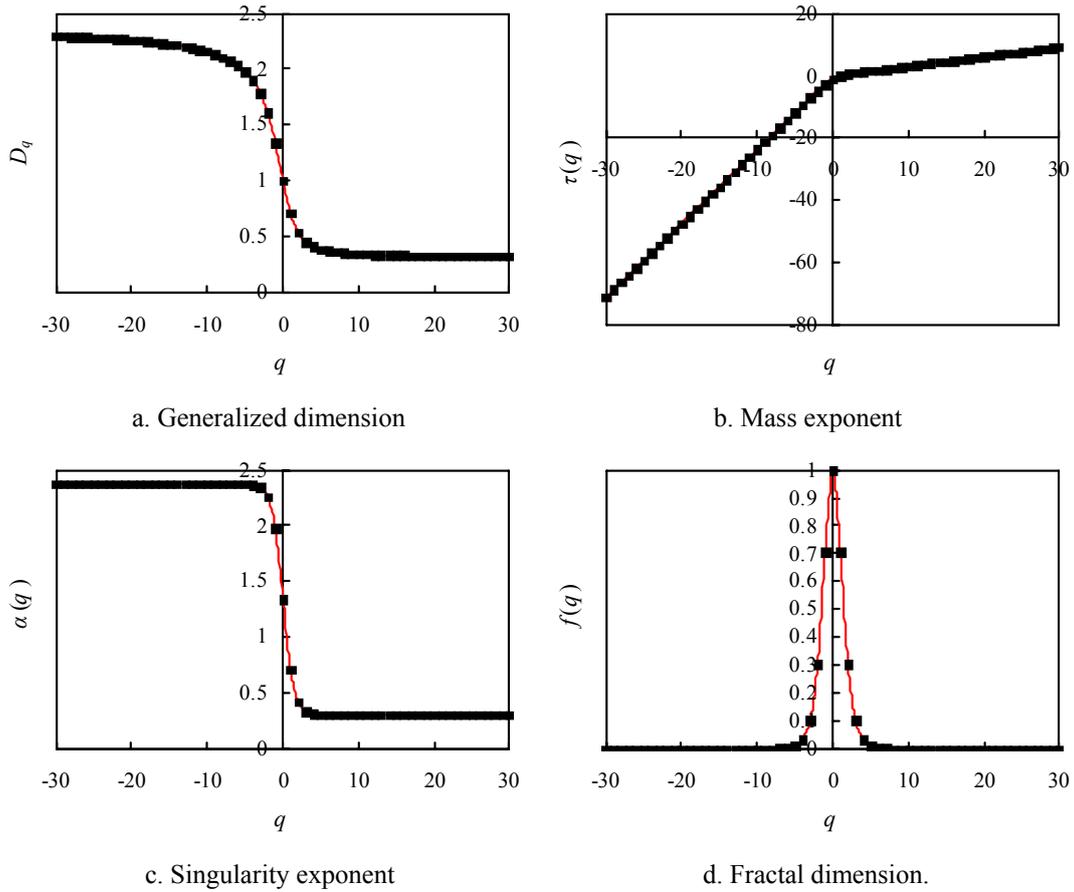

a. Generalized dimension          b. Mass exponent

c. Singularity exponent          d. Fractal dimension.

**Figure 1 The parameter spectrums of multifractal structure of the US urban-rural population**

**distribution (2010)**

[**Note**: According to the new definition of American cities, the urbanization ratio of US in 2010 is about 80.73%.]

### 3.2 Multifractal spectrums of China's urban-rural structure

The multifractal modelling can also be applied to China's urbanization. According to the sixth census data, the urbanization ratio of China is about $L$=0.4968; thus the rural population ratio is around 1-$L$=0.5032. Using the similar approach to that shown above, we can compute the multifractal parameters of Chinese urban and rural population distribution (Table 2). Based on the results, the multifractal dimension spectrums can be visually displayed (Figure 2).

Comparing the multifractal parameter spectrums of China's urban-rural structure with those of the US urban-rural structure, we can find that the ranges of fractal parameters of China's urbanization are very narrow. The maximum correlation dimension of US urban-rural structure is about $D_{-\infty}$=ln(0.1927)/ln(1/2)=2.3756, the corresponding minimum correlation dimension is about $D_{-\alpha}$=ln(0.8073)/ln(1/2)=0.3088. The maximum value depends on the ratio of urban population to total population. However, for China, the situation is different. The maximum correlation dimension of US urban-rural structure is about $D_{-\alpha}$=ln(0.4968)/ln(1/2)=1.0093, the corresponding minimum correlation dimension is about $D_{\alpha}$=ln(0.5032)/ln(1/2)=0.9908. The maximum value depends on the ratio of rural population to total population. In fact, the relationship between the singularity exponent and the local fractal dimension yield a multifractal spectrum, which is termed "$f(\alpha)$ curve". The $f(\alpha)$ curves show the difference between the multifractal spectrum of the US urbanization and that of China's urbanization (Figure 3). For the US cities, the singularity exponent value ranges from 0.31 to 2.38; while for Chinese cities, the singularity exponent value varies 0.99 to 1.01. There is no significant different between the lower limit and the upper limit of the singularity exponent of China's urbanization. In other words, the singularity exponent and thus the corresponding generalized correlation dimension of Chinese urban-rural regions can be treated as constants. This indicates that China's urbanization in 2010 is in a critical state.

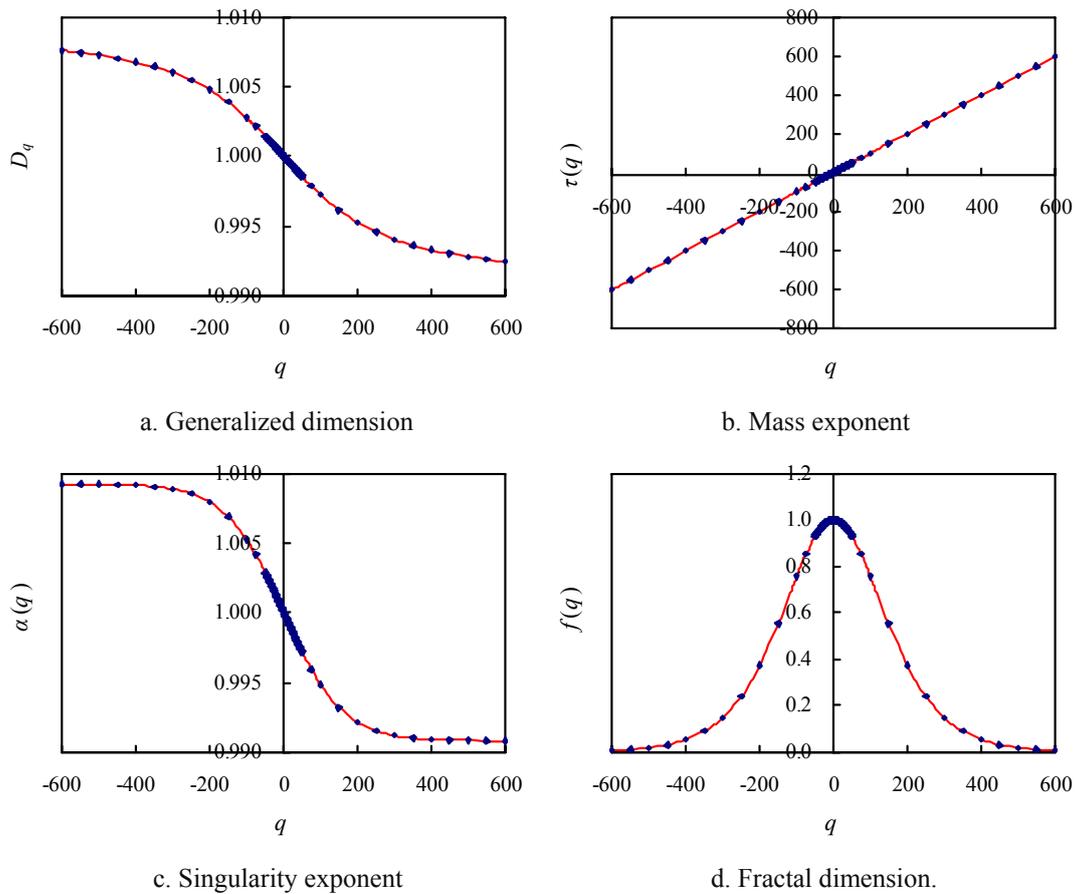

a. Generalized dimension       b. Mass exponent

c. Singularity exponent       d. Fractal dimension.

**Figure 2 The parameter spectrums of multifractal structure of China's urban-rural population distribution (2010)**

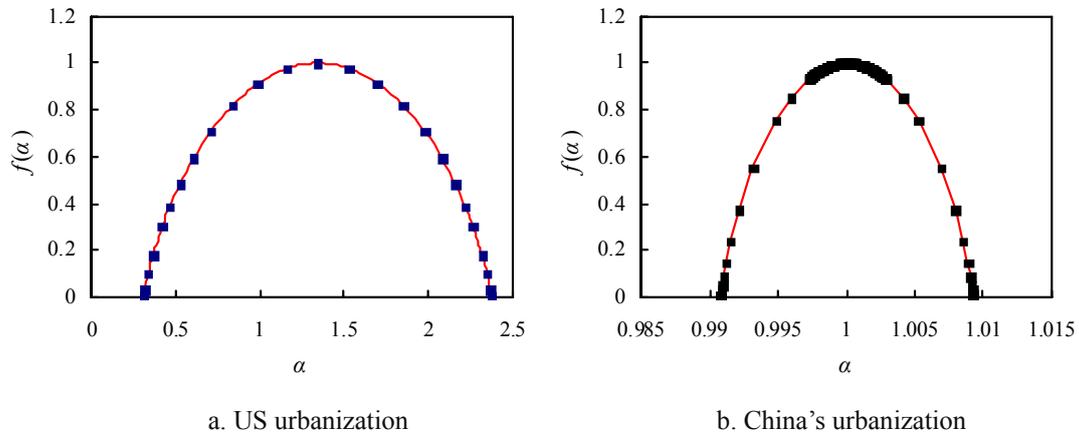

a. US urbanization    b. China's urbanization

**Figure 3 The singularity spectrums of the multifractal structure of the US and China's urban-rural population distributions (2010)**

# 4 Questions and discussion

The studies of social science fall into three types: behavioral study, axiological study, and canonical study (Krone, 1980). The behavioral study is to reveal the real patterns and processes of system development, the canonical study is to find the ideal or optimum patterns and processes for system design, and the axiological study is to construct the evaluation criterions for merits and demerits, success or failure, advantages and disadvantages, and so on. For geography, the behavioral study belongs to positive geography, the canonical study belongs to normative geography, and the axiological study can be used to connect the behavioral study and the canonical study. Fractal geometry used to be employed to make positive studies on cities. This paper is not so much a positive study (behavioral study) as a normative study (canonical study). Its main deficiency lies in that the multifractal spectrums shown above are based on a theoretical approach instead of a practical approach. Actually, the aim of this study is to lay the foundation for future definition of urban and rural using the ideas from fractals. The multifractal model of urban-rural region can be developed in both theoretical and practical directions (Figure 4). In theory, it can be linked with the replacement dynamics and allometric growth, and in practice, it can be used to define urban-rural boundary and space-filling indexes.

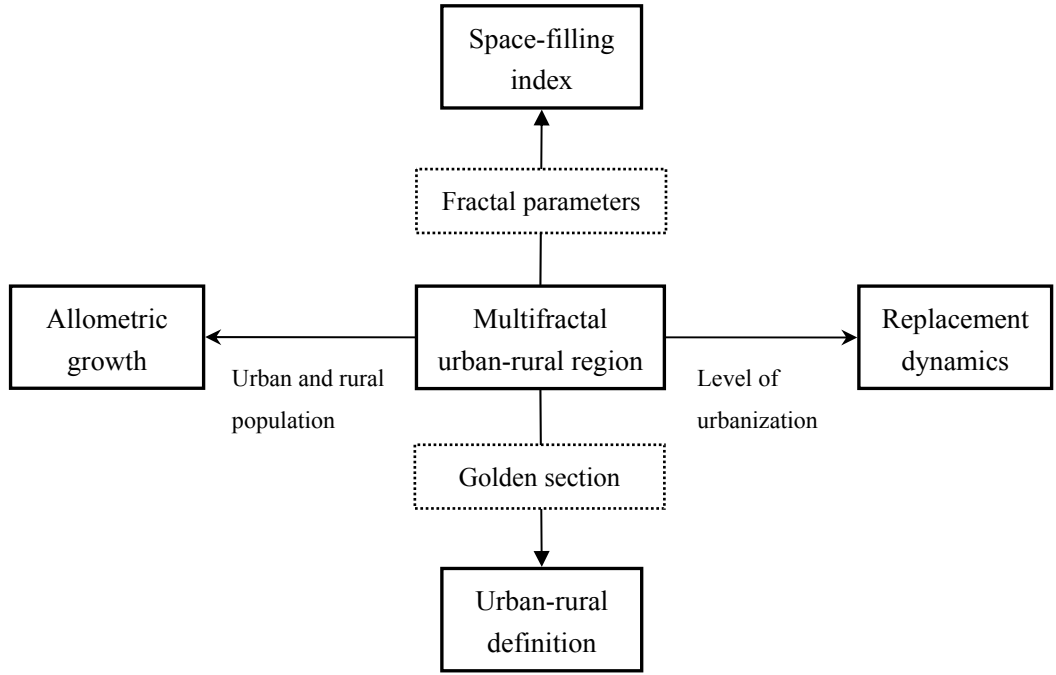

**Figure 4 The theoretical extension and method development of multifractal urban-rural region modeling**

A geographical fractal pattern is always associated with a dynamical process. Urbanization can be treated as a complex process of phase transition (Anderson *et al*, 2002; Chen, 2004; Sanders *et al*, 1997): a human geographical region evolves from the state of rural majority (the level of urbanization is less than 50%) into the state of urban majority (the level of urbanization is greater than 50%). In urban geography, a state of rural majority suggests a state of urban minority, while a state of urban majority suggests a state of rural minority (Knox and Marston, 2007). The critical state is that the ratio of urban population to total population equals that of rural population to total population. If the capacity of urbanization is $L$=1 (100%), then $L$=1/2 (50%) will indicate the critical state, and $L$=1/2 represents a threshold value of urbanization (Table 3). If $L$<1/2, we will have multifractal a pattern of rural majority. In this case, the maximum value of the generalized correlation dimension, $D_{-\infty}$, is dominated by the $L$ value; correspondingly, the minimum value of the correlation dimension, $D_{+\infty}$, is determined by the (1-$L$) value. If $L$>1/2, we will have a multifractal pattern of urban majority. In this instance, the maximum value of the generalized correlation dimension, $D_{-\infty}$, is controlled by the (1-$L$) value. Accordingly, the minimum value of the correlation dimension, $D_{+\infty}$, is determined by the $L$ value. If $L$=1/2, the multifractal pattern will reduce to a monofractal pattern, which comes between rural majority and urban majority. As for the cases shown in Section 3, the US's urban-rural distribution in 2010 displayed a multifractal state of urban majority; however, China's urban-rural structure is close to the critical state because the urbanization level of China in 2010 value has no significant difference from $L$=1/2.

**Table 3 Multifractal evolution of urbanization process: from rural majority state to urban majority state**

| Urbanization state | $L$ | 1-$L$ | $D_{-\infty}$ | $D_{+\infty}$ | $E=(D_{-\infty}-D_{+\infty})/2$ |
|---|---|---|---|---|---|
| Extreme rural state | 0 | 1 | --- | 0.0000 | --- |
| | 0.1 | 0.9 | 3.3219 | 0.1520 | 1.5850 |
| Urban minority | 0.2 | 0.8 | 2.3219 | 0.3219 | 1.0000 |
| (Rural multifractals) | 0.3 | 0.7 | 1.7370 | 0.5146 | 0.6112 |
| | 0.4 | 0.6 | 1.3219 | 0.7370 | 0.2925 |
| Critical state (monofractals) | 0.5 | 0.5 | 1.0000 | 1.0000 | 0.0000 |
| | 0.6 | 0.4 | 1.3219 | 0.7370 | 0.2925 |
| Urban majority | 0.7 | 0.3 | 1.7370 | 0.5146 | 0.6112 |
| (Urban multifractals) | 0.8 | 0.2 | 2.3219 | 0.3219 | 1.0000 |
| | 0.9 | 0.1 | 3.3219 | 0.1520 | 1.5850 |
| Extreme urban state | 1 | 0 | --- | 0.0000 | --- |

The phase transition of urbanization indicates a complex dynamics of urban-rural replacement of population (Chen, 2014c; Rao *et al*, 1989). A replacement process always takes on a sigmoid curve. The growth curve of urbanization level over time can be generally given by

$$L(t) = \frac{L_{\max}}{1 + (L_{\max} / L_0 - 1)e^{-bt^c}}, \tag{9}$$

where $L(t)$ denotes the level of urbanization of a country of time $t$, $L_0$ refers to the initial value of the urbanization level, $L_{\max}$ to the terminal value of the urbanization level (the capacity of urbanization, in theory, $L_{\max}$=1), $b$ is the initial rate of growth, and $c$ is a rate-controlling parameter, which varies from 1/2 to 2. If $c$=1, equation (9) will reduce to the logistic function, which is suitable for the developed countries (Karmeshu, 1988). If $c$=2, equation (9) will change to the quadric logistic function, which is suitable for the developing countries (Chen, 2014d). Based on the sigmoid curve of urbanization level, an urbanization process can be divided into four stages: initial stage, acceleration stage ($L(t)<L_{\max}/2$), deceleration stage ($L(t)>L_{\max}/2$), and terminal stage.

The first two stages correspond to the urbanization state of rural majority, while the last two stages correspond to the state of urban majority. The four stages of *urbanization transition* (UT) are consistent with the four phases of *demographic transition* (DT) as well as the four stages of industrialization indicative of *social transition* (ST) (Table 4). Where the macro level is concerned, the precondition of urbanization is industrialization (Knox, 2005). Demographic transition is a complex process of social evolution (Caldwell *et al*, 2006; Davis, 1945; Dudley, 1996; Notestein, 1945), which is related to urbanization dynamics.

**Table 4 Corresponding relationships between urbanization, demographic transition, and industrialization**

| Phase | Urbanization (UT) | Demographic transition (DT) | | Industrialization (ST) | Urbanization state |
|---|---|---|---|---|---|
| **First phase** | Initial stage | High phase | stationary | Agricultural society (preindustrial stage) | Urban minority |
| **Second phase** | Acceleration stage | Early phase | expanding | Early industrial society (early industrial stage) | (Rural majority) |
| **Third phase** | Deceleration stage | Late phase | expanding | Late industrial society (late industrial stage) | Urban majority |
| **Fourth phase** | Terminal stage | Low phase | stationary | Information society (postindustrial stage) | (Rural minority) |

Urbanization involves urban form of intraurban geography and urban systems of interurban geography (Knox and Marston, 2007). The patterns of urban form are associated with the process of urban growth. Thus, urbanization dynamics is correlated with the dynamics of urban growth. Corresponding to the urbanization process, urban growth takes on a sigmoid curve and can be measured with fractal dimension of urban form such as (Chen, 2014c)

$$D(t) = \frac{D_{\max}}{1 + (D_{\max}/D_0 - 1)e^{-kt^v}}, \qquad (10)$$

where $D(t)$ is the fractal dimension of a city of time $t$, $D_0$ denotes to the initial value of the fractal dimension, $D_{\max}$ to the terminal value of the fractal dimension (the capacity of fractal dimension, in theory, $D_{\max}$=2), $k$ is the original rate of growth, and $v$ is a rate-controlling parameter coming between 1 and 2. Empirically, the $v$ value of a city is always equal to the $c$ value of the corresponding urbanization model. For the cities of developed countries, $v$=$c$=1; for the cities of developing countries, $v=c=2$. Therefore, the process of urban growth can also be divided into four stages: initial stage, acceleration stage ($D(t)<D_{max}/2$), deceleration stage ($D(t)>D_{max}/2$), and terminal stage.

The multifractal parameters can be related to the allometric scaling and spatial dynamics of urbanization. At the initial and acceleration stages of urbanization, the relationships between urban population and rural population often follow the law of allometric growth (Naroll and Bertalanffy, 1956), which can be expressed as

$$u(t) = ar(t)^b,\qquad(11)$$

where $t$ denotes time, $a$ refers to the proportionality coefficient, and $b$ to the scaling exponent. Accordingly, the growth of urbanization level over time is as below:

$$L(t) = \frac{u(t)}{u(t)+r(t)} = \frac{ar(t)^b}{ar(t)^b+r(t)} = \frac{ar(t)^{b-1}}{1+ar(t)^{b-1}}.\qquad(12)$$

Thus the global parameters of urbanization multifractals can be expressed as

$$D_q(t) = \frac{\tau(q,t)}{q-1} = \frac{\ln[L(t)^q+(1-L(t))^q]}{(1-q)\ln 2} = \frac{\ln[(\frac{ar(t)^{b-1}}{1+ar(t)^{b-1}})^q+(\frac{1}{1+ar(t)^{b-1}})^q]}{(1-q)\ln 2}.\qquad(13)$$

Further, the local parameters of urbanization multifractals can be derived from equation (13) by means of Legendre's transform.

The multifractal parameters can be used to measure the degree of geo-spatial utilization in terms of urbanization. Based on the generalized correlation dimension, a space-filling index can be defined as below:

$$E = \frac{D_{-\infty}-D_{+\infty}}{d} = \frac{\max(D_q)-\min(D_q)}{2},\qquad(14)$$

where $E$ denotes the space-filling index. If $L<L_{max}/2$, the index indicates the rural geo-spatial utilization efficiency, reflecting the rural space-filling extent; If $L>L_{max}/2$, the index implies the urban geo-spatial utilization efficiency, reflecting the urban space-filling extent. The space-filling index can also be defined based on the singularity exponent, and the formula is

$$E = \frac{\alpha(-\infty)-\alpha(+\infty)}{d} = \frac{\max[\alpha(q)]-\min[\alpha(q)]}{2}.\qquad(15)$$

Because of urbanization, the rural space-filling index goes down and down, while the urban space-filling index goes up and up. Using the census data of urbanization, we can estimate the urban and rural space-filling indexes of America and China in different years (Tables 5 and 6).

**Table 5 The US urban and rural space-filling indexes based on urbanization level (1790-2010)**

| Urbanization state | Year | $L$ | $1-L$ | $D_{-\infty}$ | $D_{+\infty}$ | $E=(D_{-\infty}-D_{+\infty})/2$ |
|---|---|---|---|---|---|---|
| **Rural majority (urban minority)** | 1790 | 0.0513 | 0.9487 | 4.2843 | 0.0760 | 2.1041 |
| | 1800 | 0.0607 | 0.9393 | 4.0415 | 0.0904 | 1.9756 |
| | 1810 | 0.0726 | 0.9274 | 3.7843 | 0.1087 | 1.8378 |
| | 1820 | 0.0719 | 0.9281 | 3.7973 | 0.1077 | 1.8448 |
| | 1830 | 0.0877 | 0.9123 | 3.5121 | 0.1323 | 1.6899 |
| | 1840 | 0.1081 | 0.8919 | 3.2092 | 0.1651 | 1.5220 |
| | 1850 | 0.1541 | 0.8459 | 2.6978 | 0.2415 | 1.2282 |
| | 1860 | 0.1977 | 0.8023 | 2.3386 | 0.3178 | 1.0104 |
| | 1870 | 0.2568 | 0.7432 | 1.9612 | 0.4282 | 0.7665 |
| | 1880 | 0.2815 | 0.7185 | 1.8286 | 0.4770 | 0.6758 |
| | 1890 | 0.3510 | 0.6490 | 1.5104 | 0.6237 | 0.4434 |
| | 1900 | 0.3965 | 0.6035 | 1.3348 | 0.7285 | 0.3031 |
| | 1910 | 0.4561 | 0.5439 | 1.1326 | 0.8785 | 0.1270 |
| **Urban majority (rural minority)** | 1920 | 0.5117 | 0.4883 | 1.0342 | 0.9666 | 0.0338 |
| | 1930 | 0.5614 | 0.4386 | 1.1889 | 0.8330 | 0.1779 |
| | 1940 | 0.5652 | 0.4348 | 1.2017 | 0.8231 | 0.1893 |
| | 1950 | 0.6400 | 0.3600 | 1.4739 | 0.6439 | 0.4150 |
| | 1960 | 0.6986 | 0.3014 | 1.7302 | 0.5175 | 0.6064 |
| | 1970 | 0.7364 | 0.2636 | 1.9236 | 0.4414 | 0.7411 |
| | 1980 | 0.7374 | 0.2626 | 1.9290 | 0.4395 | 0.7447 |
| | 1990 | 0.7521 | 0.2479 | 2.0121 | 0.4110 | 0.8006 |
| | 2000 | 0.7901 | 0.2099 | 2.2524 | 0.3398 | 0.9563 |
| | 2010 | 0.8073 | 0.1927 | 2.3756 | 0.3088 | 1.0334 |

**Note:** The original data of the US level of urbanization are available from the US Census Bureau's website: http://www.census.gov/population.

**Table 6 China's rural space-filling indexes based on urbanization level (1953-2010)**

| Urbanization state | Year | $L$ | $1-L$ | $D_{-\infty}$ | $D_{+\infty}$ | $E=(D_{-\infty}-D_{+\infty})/2$ |
|---|---|---|---|---|---|---|
| **Rural majority (urban minority)** | 1953 | 0.1326 | 0.8674 | 2.9148 | 0.2052 | 1.3548 |
| | 1964 | 0.1410 | 0.8590 | 2.8262 | 0.2193 | 1.3035 |
| | 1982 | 0.2055 | 0.7945 | 2.2828 | 0.3319 | 0.9755 |
| | 1990 | 0.2623 | 0.7377 | 1.9307 | 0.4389 | 0.7459 |
| | 2000 | 0.3609 | 0.6391 | 1.4703 | 0.6459 | 0.4122 |
| | 2010 | 0.4968 | 0.5032 | 1.0093 | 0.9908 | 0.0092 |

**Note:** The original data of the US level of urbanization are available from the website of National Bureau of Statistics of the People's Republic of China: http://www.stats.gov.cn/tjsj/ndsj/.

Urban form has no characteristic scale, and thus an urban boundary cannot be identified exactly. The concept of city is actually based on subjective definitions rather than objective measurements. In this case, the golden section can be employed to optimize the definition of urban-rural regions. In fact, there exist two basic measurements for urbanization. One is the level of urbanization, and the other, urban-rural ratio (United Nations, 1980; United Nations, 2004). The relation between the two measurements is as below:

$$L = \frac{u}{u+r} = \frac{1}{1+r/u} = \frac{1}{1+1/O},$$  (16)

where $O=u/r$ denotes the urban-rural ratio. Suppose that the ideal state of the terminal stage is the level of urbanization equals the rural-urban ratio, that is, $L=1/O=r/u$. Thus we have $L=1/(1+L)$, from which it follows

$$L^2 + L - 1 = 0.$$  (17)

One of solutions of the quadratic equation is

$$L = \frac{-1 + \sqrt{1 - 4 \times (-1)}}{2} = \frac{\sqrt{5} - 1}{2} \approx 0.618,$$

which is just the golden ratio. The corresponding urban-rural ratio is $O=1/L\approx1.618$, which is also termed golden ratio. Letting $L=0.618$ and $1-L=0.382$, we can obtain multifractal parameter spectrums based on the golden mean. This suggests a possibility that we can define the urban-rural regions according to the golden ratio. Of course, this is just a speculation at present.

## 5 Conclusions

Human geographical systems differ from the classical physical systems because that the laws of human geography are not of spatio-temporal translational symmetry. Geographical studies are significantly different from physical studies. Physics focuses on only facts and cause-and-effect behavioral relationships in the real world. However, human geography involves both observational facts in the real world and value or normative judgments in the possible world or the ideal world. This paper focuses on value judgments of spatial structure of human geographical systems rather than facts and causality of urban and rural behaviors. Based on the theoretical analysis and empirical evidences, the main conclusions can be reached as follows.

**First, multifractal measures can be employed to characterize or define urban-rural geographical patterns.** If a human geographical system in the real world is of multi-scaling fractal structure, multifractal geometry can be used to characterize the urban-rural terrain systems and make empirical analyses of urban evolution; if a real human geographical system is not of multifractals, multifractal theory can be used to optimize the urban-rural spatial structure. Multifractality represents optimal structure of human geographical systems because a fractal object can occupy its space in the most efficient way. Using the ideas from multifractals to design or plan urban and rural terrain systems, we can make the best of human geographical space.

**Second, multifractal parameters can be adopted to model the dynamical process of urban and rural evolution**. Urbanization is a complex process of urban-rural replacement, which is associated with critical phase transition: from the state of rural majority (urban minority) to the state of urban majority (rural minority). There is critical state coming between the rural majority and urban majority. The phase of rural majority corresponds to a rural multifractal pattern, while the phase of urban majority corresponds to an urban multifractal pattern. In theory, the critical state corresponds to a transitory monofractal pattern. Moreover, the multifractal urban-rural model can b associated with allometric growth, which indicates spatial scaling and nonlinear dynamics.

**Third, the generalized correlation dimension and the singularity exponent can be used to define a space-filling index based on the level of urbanization**. This space-filling index makes a measurement of geo-spatial utilization, and it can be termed urban-rural utilization coefficient. If the level of urbanization is less than 1/2, the index implies the rural geo-spatial utilization coefficient indicating the rural space-filling degree; if the level of urbanization is greater than 1/2, the index denotes the urban geo-spatial utilization coefficient indicative of the urban space-filling degree. Along with urbanization, the rural space-filling index descends, while the urban space-filling index ascends gradually. A conjecture is that the golden section and multifractal ideas can be combined to define urban boundaries and thus optimize urban-rural patterns.

## Acknowledgment

This research was sponsored by the National Natural Science Foundation of China (Grant No.

41171129). The supports are gratefully acknowledged.